\documentclass[aip,amsmath,amssymb,reprint]{revtex4-2}
\usepackage{graphicx}
\usepackage[utf8]{inputenc}
\usepackage[T1]{fontenc}
\usepackage[pdfstartview=FitH, pdfpagemode=None, pdfpagelayout=OneColumn, colorlinks=true, linkcolor=blue, citecolor = blue]{hyperref}

\begin{document}

\title{Oscillation damper for misaligned witness in plasma wakefield accelerator}

\author{Konstantin V. Lotov, Ivan Yu. Kargapolov, Petr V. Tuev}
\affiliation{Novosibirsk State University, Novosibirsk 630090, Russia}
\affiliation{Budker Institute of Nuclear Physics, Novosibirsk 630090, Russia}
\date{\today}
\begin{abstract}
If a laser- or particle beam-driven plasma wakefield accelerator operates in the linear or moderately nonlinear regime, injecting an externally produced particle bunch (witness) to be accelerated may encounter an alignment problem.
Witness alignment tolerances can be relaxed by using a damper, an additional particle bunch produced by the same injector and propagating at a submillimeter distance ahead of the witness.
If misaligned, the damper perturbs the wakefield in such a way that the witness shifts on-axis with no quality loss.
\end{abstract}

\maketitle

Wakefield acceleration of particles in plasma, in which a strong field is excited in the plasma by a compact laser pulse or a relativistic particle bunch, promises a drastic reduction in the size of high-energy particle accelerators.\cite{PoP27-070602, NJP23-031101, RMP90-035002}
The quality of the accelerated beam (called witness) in this method depends largely on the injection of particles into the plasma wave.\cite{PoP19-055501, AIP1507-67, RMP81-1229}
Most injection techniques rely on trapping of plasma electrons by the wave.
Injection of an externally produced beam from a conventional accelerator is also of interest,\cite{PScr-T30-110, PRL68-48, PoP1-1753, JAP76-7645, PRL74-5220, AIP395-343, PRL81-995, PRL92-095004, AIP1507-722, NIMA-740-60, NIMA-829-229, NIMA-909-46, PPCF60-034001, EPJST229-3675, Symmetry14-1680} as it offers additional degrees of freedom in shaping and positioning of the witness, and in some cases is the only applicable method.
Our study relates to the latter option.

\begin{figure}[b]
\includegraphics{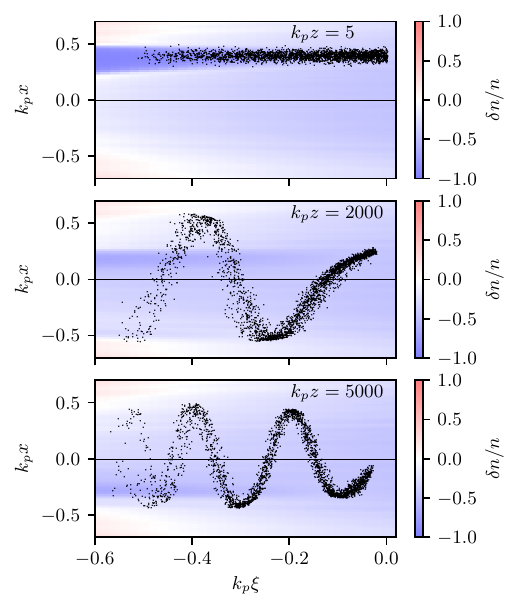}
\caption{Distortion of a misaligned electron witness as it propagates through the plasma. 
The black points show the witness particles, the color shows the perturbation $\delta n$ of the plasma electron density. 
The witness propagates to the right (in the $z$-direction). }
\label{fig1-waving}
\end{figure}

The external witness injection comes along with the alignment problem.
The witness and wakefield axes must coincide to a small fraction of $k_p^{-1}$, where $k_p = \sqrt{4 \pi n e^2 / (mc^2)}$ is the plasma wavenumber, $n$ is the plasma density, $m$ is the electron mass, $e$ is the elementary charge, and $c$ is the speed of light.
Otherwise, the witness wriggles around the wakefield axis (Fig.\,\ref{fig1-waving}) because the focusing force in the plasma depends on the co-moving longitudinal coordinate $\xi = z-ct$ and different parts of the  witness oscillate transversely with different frequencies.
As a result, the quality of the witness deteriorates, i.e. its radius and emittance increase.
This was not a problem in early proof-of-principle experiments with external injection,\cite{PScr-T30-110, PRL68-48, PoP1-1753, JAP76-7645, PRL74-5220, AIP395-343, PRL81-995, PRL92-095004} but it may be an issue in future experiments aimed at demonstrating witness quality.\cite{AIP1507-722, NIMA-740-60, NIMA-829-229, NIMA-909-46, PPCF60-034001, EPJST229-3675, Symmetry14-1680}
The required tolerances may be difficult to achieve.\cite{PRAB21-011301, PRAB25-101602, NIMA-1049-168094} 
Only the blowout regime of acceleration\cite{APB74-355, PRA44-6189} is free from this source of emittance growth, but this regime usually involves the injection of plasma electrons into the wave.

In this paper, we propose a method which can relax requirements on witness-to-driver alignment.
The method works for linear or moderately nonlinear wakefields only, but this is exactly where it is most needed.

\begin{figure}[tb]
\includegraphics{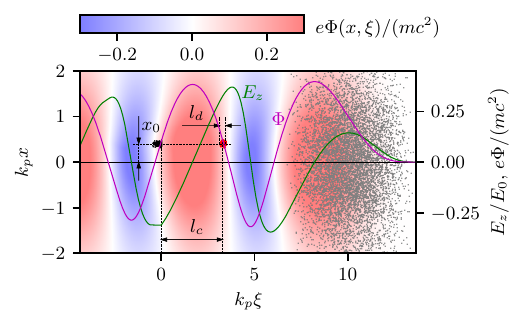}
\caption{Illustration of the damper idea. 
The driver, witness, and damper are shown in gray, black, and red points, respectively. 
The colored background is the wakefield potential $\Phi(x,\xi)$. 
The overlaid lines are the electric field $E_z$ and wakefield potential $\Phi$ on the axis as if there is no damper and the witness is exactly on the axis. }
\label{fig2-idea}
\end{figure}
\begin{figure}[tb]
\includegraphics{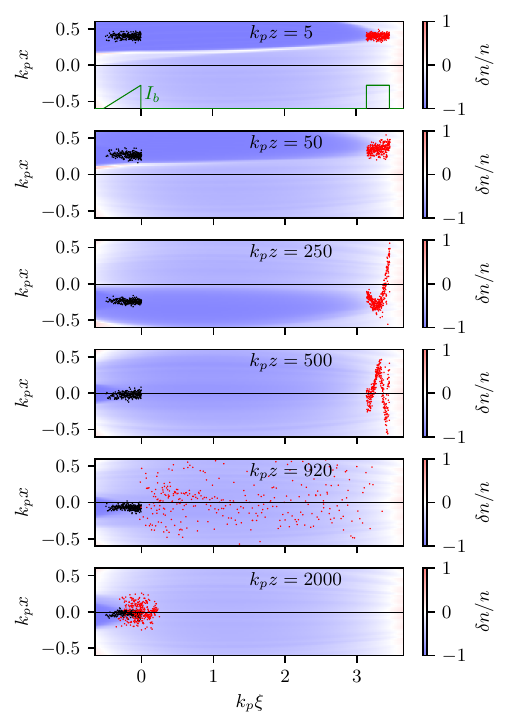}
\caption{Evolution of misaligned electron witness (black points) and damper (red points) as they propagate in the plasma.
The colored background is the perturbation of plasma electron density $\delta n$.
The green line in the upper fragment schematically shows the current $I_b$ of the beams.}
\label{fig3-damper}
\end{figure}

The idea is illustrated by Fig.\,\ref{fig2-idea}.
The witness is preceded by an additional bunch of particles of the same sort, which we call a damper.
It is assumed that the damper is produced by the same injector as the witness. 
If the witness is misaligned with the driver axis, the damper is misaligned by the same amount $x_0$. 
The damper is in approximately the same focusing field as the witness.
If misaligned, the damper oscillates transversely with a frequency close to the witness oscillation frequency, and the wakefield perturbation induced by the damper dampens the witness oscillations (Fig.\,\ref{fig3-damper}). 
As a result, the witness aligns to the axis without quality loss.

This method relies on several physical principles and technological advances.

First, there is always a cross-section in which the frequency of damper oscillations equals that of the witness head. 
The electromagnetic force $\vec F$ acting on the witness or damper particles is proportional to the gradient of some scalar function $\Phi$, which we call the wakefield potential: $\vec F = -q \nabla \Phi$, where $q$ is the particle charge. 
The witness sits on the rear slope of the potential well, where it is both focused and accelerated by the wave. 
On the opposite side of the potential well, there is always a point where the focusing force is the same and the longitudinal field is decelerating (Fig.\,\ref{fig2-idea}).

Second, the technique of generating electron bunch trains or bunches of a given shape is now rapidly developing.\cite{RMP94-025006, JINST17-P05036} 
Producing a pair of bunches with a controlled sub-millimeter spacing is within the state of the art.
It is not always possible to make all bunches of the highest quality and exactly the same energy, but this is not necessary. 
The damper could have a different energy, larger emittance, or larger radius than the witness and it would still work well.

Third, if there are plasma electrons around the damper (as it is if the blowout is not reached), the damper locally perturbs the wave, and this perturbation is transferred by the wave exactly to the witness position (Fig.\,\ref{fig3-damper}).

Fourth, the damper's effect on the witness is time-limited. 
There are two reasons for this: damper deceleration and phase mixing of damper oscillations. 
Because of the limited action, the damper can ``place'' the witness onto the axis and ``leave'' it in this position. 

Fifth, the damper gains a lower energy than the witness and can subsequently be separated in energy.
When the damper loses energy, it expands and shifts backward into the accelerating phase (Fig.\,\ref{fig3-damper}). 
Since it was initially at approximately the same wakefield potential as the witness, it shifts to the same position where the witness is.
The damper loads the wave and must be taken into account when matching the witness and the wave.
However, the damper energy is lower than the witness energy by about twice the initial damper energy, as the witness gains energy while the damper slows down by a field of the same amplitude.

Sixth, and the most important, in a moderately nonlinear axially symmetric plasma wave, small transverse oscillations of beam particles are harmonic. 
The oscillation frequency does not depend on the magnitude and direction of the offset if the magnitude is much smaller than the characteristic transverse scale of the plasma wave $k_p^{-1}$. 
The ``golden ratio'' of oscillation frequencies (damper-to-witness) at which the witness is placed on the axis, found for some initial offset $x_0 \ll k_p^{-1}$, is the same for other offsets. 
Therefore, for a certain damper charge and witness-to-damper distance, the witness comes onto the axis for any transverse offset in any direction, if this offset is not too large.

\begin{table}[tb]
\begin{center}
\caption{Parameters of the illustrative variant}\label{t1}
\begin{tabular}{llc}\hline\hline
  Parameter, notation && Value \\ \hline
  Driver radius, $\sigma_{rd}$ && $k_p^{-1}$ \\
  Driver length, $\sigma_{zd}$ && $1.05 k_p^{-1}$ \\
  Peak driver density, $n_d$ && $0.4n$ \\
  Witness and damper radius, $\sigma_r$ && $0.03 k_p^{-1}$ \\
  Witness and damper peak current, $I_w$ && $0.0268 mc^3 / e$ \\
  Witness and damper normalized emittance && $0.01 k_p^{-1}$ \\
  Witness length, $l_w$ && $0.52 k_p^{-1}$ \\
  Damper length, $l_d$ && $0.32 k_p^{-1}$ \\
  Witness-to-damper distance, $l_c$ && $0.36 k_p^{-1}$ \\
  Witness and damper energy, $W_0$ && 150\,MeV \\
  Witness and damper offset, $x_0$ && $0.4 k_p^{-1}$ \\
  Grid steps in $x$, $y$, $\xi$ && $0.02 k_p^{-1}$ \\
  Step in $z$ for updating plasma state && $5 k_p^{-1}$ \\
  Number of electrons per cell && 9 \\
  Simulation window width in $x$ and $y$ && $8 k_p^{-1}$ \\ \hline
  \hline
 \end{tabular}\end{center}
\end{table}

Let us numerically demonstrate the damper operation at parameters close to those at which the external injection could be used (Table~\ref{t1}).
All figures in this paper are made for this parameter set unless stated otherwise.
The simulations are performed with the three-dimensional quasistatic code LCODE.\cite{noise}

In this example, a wave with maximum field amplitude $E_m \approx 0.35 E_0$ is excited by a single non-evolving electron bunch. 
The witness charge, shape, and location are chosen so that all witness particles are in the accelerating field $E_\text{acc} \approx 0.9 E_m \approx 0.311 E_0$ if there is no damper and the witness is exactly on the axis. 
The witness has a triangular current profile that minimizes the energy spread\cite{PAcc22-81} (Fig.\,\ref{fig3-damper}). 
The damper has a constant current equal to the maximum witness current, so the damper length $l_d$ also determines its charge.
The witness and damper have the same radius and emittance, matched to the focusing force of the ion background.\cite{PoP2-1326, PRAB21-011301}
Since the density of both exceeds that of the plasma, the witness and damper create their own bubbles, if they not expand due to wriggling.

For the plasma density $n = 7 \times 10^{14}\,\text{cm}^{-3}$, this parameter set corresponds to acceleration of 80\,pC electron bunch of radius $6\,\mu$m, total length $100\,\mu$m and normalized emittance 2.2\,mm\,mrad at the rate of 0.8\,GeV/m. 
This is close to the parameters of electron injection for AWAKE.\cite{PRAB21-011301, JPCS1596-012008, Symmetry14-1680} 
However, in our example, the witness is accelerated by a field twice as strong as in Ref.~\onlinecite{PRAB21-011301} and oscillates at three times the frequency if misaligned, so the focusing force of the wave is not as small compared to that of the pure ion background as in Ref.~\onlinecite{PRAB21-011301}. 
As a consequence, propagating in its own bubble does not keep the witness from wrigging.


For the plasma density $n = 10^{17}\,\text{cm}^{-3}$, our parameters correspond to acceleration of 7\,pC electron bunch of radius $0.5\,\mu$m, length $9\,\mu$m and normalized emittance 0.18\,mm\,mrad at the rate of 9\,GeV/m. 
This is within the parameter range covered by several projects,\cite{NIMA-740-60, NIMA-829-229, EPJST229-3675, PRAB23-031301} but does not match any of them exactly.

\begin{figure}[tb]
\includegraphics{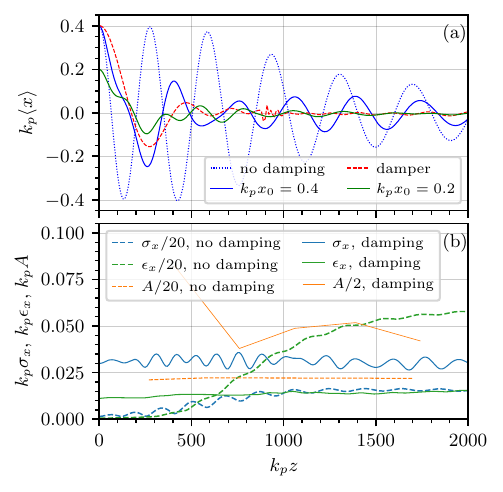}
\caption{(a) Transverse oscillations of witness or damper center of mass $\langle x \rangle$ with or without damping, also for the reduced initial offset $x_0 = 0.2 k_p^{-1}$; (b) evolution of witness rms size $\sigma_x$, normalized emittance $\epsilon_x$ and effective size $A$ with and without damping. Some graphs are shrunk vertically to fit fragment (b).}
\label{fig4-lines}
\end{figure}

Without a damper, the witness oscillates about the axis, its root-mean-square (rms) width in the offset plane increases to roughly half of the initial offset, and the emittance increases accordingly (Fig.\,\ref{fig4-lines}). 
An optimally matched damper oscillates at a lower frequency and slows the transverse motion of the witness. 
It takes a few oscillation periods for the damper to stop witness oscillations. 
In contrast to idealized expectations, the similarity of witness and damper oscillations at different offsets [$k_p x_0 = 0.4$ and $k_p x_0 = 0.2$ in Fig.\,\ref{fig4-lines}(a)] holds only approximately and mostly during the first period, but this is sufficient for damping.

\begin{figure}[tb]
\includegraphics{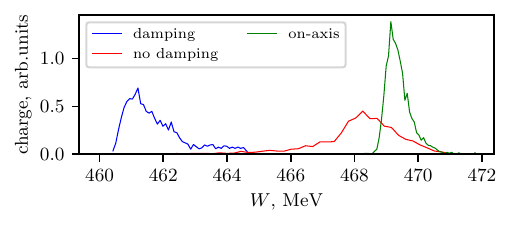}
\caption{ Energy spectra at $k_p z = 2000$ of the initially misaligned witness with and without the damper, and of the witness injected perfectly on-axis.}
\label{fig5-spectra}
\end{figure}

In the example considered, the witness is aligned with respect to the axis so that the amplitude of its residual oscillations does not exceed $x_0/5$, its radius remains initial, and its emittance increases by only 30\% (Fig.\,\ref{fig4-lines}). 
Without the damper, the emittance increases by a factor of 60 and the rms beam width exceeds half the initial offset.
The effect of the damper on the witness energy spread is comparable to that of the misalignment (Fig.\,\ref{fig5-spectra}) and can be eliminated by appropriate modification of the witness current profile. 
At the same moment (at $k_p z = 2000$) the damper particles have energies from 140 and 167\,MeV, so the damper does not overlap with the witness in energy.

The question arises as to how wide the damping regime is in the parameter space. 
To answer, let us vary three main parameters that control the process: damper location $l_c$, length $l_d$ and offset $x_0$. 
We also need a target function to compare the variants. 
As such, let us take the quantity
\begin{equation}\label{we1}
    A = k_p \sqrt{ \langle x \rangle_m^2 + \sigma_x^2 } \left( \frac{\langle p_z \rangle}{p_{z0}} \right)^{1/4},
\end{equation} 
where $\langle x \rangle_m$ is the amplitude of witness oscillations [local maxima of the function $\langle x \rangle (z)$ shown in Fig.\,\ref{fig4-lines}(a)], $\sigma_x$ is the rms witness width [shown in Fig.\,\ref{fig4-lines}(b)], $\langle p_z \rangle$ is the average $z$-momentum of witness particles, and $p_{z0}$ is the initial witness momentum. 
The function (\ref{we1}) is convenient in that it quickly approaches a constant and does not change further as the witness accelerates [Fig.\,\ref{fig4-lines}(b)]. 
In simulations, we measure it at $k_p z = 2000$.
We call it the effective witness size.

\begin{figure}[tb]
\includegraphics{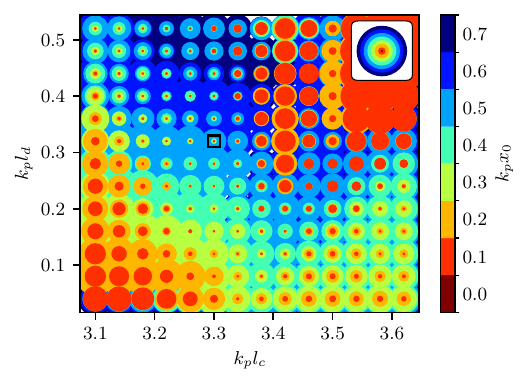}
\caption{ Map of the effective witness sizes represented by circles of radius $A$. 
The color of the circles shows the initial offset $x_0$.
The upper right inset shows the values of $A$ without damping, taken at $z=0$.
The small black square marks the variant chosen for the illustrations.}
\label{fig6-map}
\end{figure}

To visualize the damping efficiency characterized by points in the four-dimensional space $(l_c, l_d, x_0, A)$, we plot circles of radius $A$ on the plane $(l_c, l_d)$, smaller offsets over larger ones (Fig.\,\ref{fig6-map}).
These circles can be imagined as witness sizes established after phase mixing of the transverse oscillations.
For ease of comparison, the inset shows the witness sizes without damping. 
If a circle is smaller than the circle of the same color on the inset, then the damper reduces witness oscillations.
In such a representation, the dominant colors of parts of the figure show the maximum offset up to which the damping works efficiently.

We see that the damping reduces the effective witness size by almost an order of magnitude over a wide parameter range.
Damping at larger offsets requires longer dampers with higher charges.
If the damper charge is higher than the witness charge (here the charges are equal at $k_p l_d = 0.26$), the damping can be effective even for initial offsets larger than $0.5 k_p^{-1}$ (in the region of Fig.\,\ref{fig6-map} where the background color is blue).
The size of this region indicates the required accuracy of damper positioning ($\sim 0.2 k_p^{-1}$) and charge ($\pm 20$\%).
It is no stricter than the tolerances required to accelerate a witness with a percent-level energy spread.

To summarize, our method can relax the tolerances on witness alignment by more than an order of magnitude if there is freedom in shaping the injected beams before entering the plasma. 
With an additional damper bunch properly positioned ahead of the witness bunch, it is sufficient to ``hit'' a circle of radius $\sim k_p^{-1}$ to accelerate the witness without loss of quality due to misalignment. 

\acknowledgements

This study was supported by the Russian Science Foundation, Project No. 23-12-00028.

\end{document}